\documentclass[article,aps,epsf,amsfonts,amssymb,amsmath,nofootinbib,eqsecnum,]{revtex4} 

\usepackage{amsmath,amssymb,amsfonts,amsthm,bbm}
\usepackage{graphicx}

\usepackage{subeqnarray}

\newcommand{\be}{\begin{equation}}\newcommand{\ee}{\end{equation}}
\newcommand{\bea}{\begin{eqnarray} }\newcommand{\eea}{\end{eqnarray}}
\newcommand{\beaa}{\begin{eqnarray} }\newcommand{\eeaa}{\end{eqnarray}}
\newcommand{\bsa}{\begin{subeqnarray}}\newcommand{\esa}{\end{subeqnarray}}
\newcommand{\ba}{\begin{array}}\newcommand{\ea}{\end{array}}
\newcommand{\bit}{\begin{itemize}}\newcommand{\eit}{\end{itemize}}
\newcommand{\ben}{\begin{enumerate}}\newcommand{\een}{\end{enumerate}}

\def\lab{\label}

\def\rar{\rightarrow}

\def\al{\alpha}\def\ga{\gamma}
\def\de{\delta}\def\ep{\epsilon}
\def\la{\lambda}
\def\si{\sigma}\def\om{\omega}
\def\Om{\Omega}%

\def\1{{_{1}}}\def\2{{_{2}}}

\def\lsim{\hbox{ \raise.35ex\rlap{$<$}\lower.6ex\hbox{$\sim$}\ }}
\def\gsim{\hbox{ \raise.35ex\rlap{$>$}\lower.6ex\hbox{$\sim$}\ }}

\def\lab{\label}

\def\rar{\rightarrow}

\def\al{\alpha}

\def\ga{\gamma}

\def\de{\delta}

\def\ep{\epsilon}

\def\la{\lambda}

\def\si{\sigma}

\def\om{\omega}
\def\Om{\Omega}
\begin{document}
\title{Self-similarity properties of nafionized and filtered water and
deformed coherent states
}

\author{A. Capolupo$^a$\footnote{capolupo@sa.infn.it}, E. Del Giudice$^b$\footnote{delgiudice@mi.infn.it}, V. Elia$^c$\footnote{vittorio.elia@unina.it}, R. Germano$^d$\footnote{germano@promete.it}, E. Napoli$^c$\footnote{elena.napoli@unina.it}, M. Niccoli$^c$\footnote{marcella.niccoli@unina.it}, A. Tedeschi$^e$\footnote{gowhite@usa.net}, G. Vitiello$^f$\footnote{vitiello@sa.infn.it}
}

\affiliation{$^a$Dipartimento di Ingegneria Industriale, Universit\'a di Salerno, I-84084 Fisciano (Salerno), Italy\\
$^b$Sezione INFN, I-20122 Milano, Italy (retired) and Centro Studi Eva Reich, Via Orti, 5, I-20122 Milano, Italy\\
$^c$Dipartimento di Scienze Chimiche, Universit\'a di Napoli ``Federico II'', I-80100 Napoli, Italy\\
$^d$PROMETE Srl, CNR Spin off, via Buongiovanni,~49,
I-80046 San Giorgio a Cremano (Napoli), Italy\\
$^e$WHITE Holographic Bioresonance, Via F. Petrarca, 16, I-20123 Milano, Italy\\
$^f$Dipartimento di Fisica ``E. R. Caianiello'' and Istituto Nazionale di Fisica Nucleare\\ Universit\'a di Salerno, I-84084 Fisciano (Salerno), Italy
}


\begin{abstract}

By resorting to measurements of physically characterizing observables of water samples perturbed by the presence of Nafion and by iterative filtration processes, we discuss their scale free, self-similar fractal properties. By use of algebraic methods the isomorphism is proved between such self-similarity features and the deformed coherent state formalism.

\vspace{5mm}


\end{abstract}

\maketitle


\section{Introduction}\label{sec1}

It has been shown recently~\cite{PLA2012,NewMat2008,QI2009,Licata2008} that in the framework of the theory of entire analytical functions
the algebra of deformed (squeezed) coherent states provides the functional realization of self-similarity properties of deterministic fractals. Such a result has been obtained by using the quantum field theory (QFT) formalism which describes topologically non-trivial ``extended objects'', such as kinks, vortices, monopoles, crystal dislocations, domain walls, and other so-called ``defects'', or {\it macroscopic quantum systems}, in condensed matter physics in terms of non-homogeneous boson condensation~\cite{Umezawa:1982nv,DifettiBook,Bunkov} and which has been tested to be successful in explaining many experimental observations in superconductors, crystals, ferromagnets, etc.~\cite{Bunkov}. The  observation~\cite{CrystalFractals} of defects in the lattice of crystals (dislocations) submitted to stress actions, such as the bending of the crystal at low temperature, may also be framed in the QFT description of defect formation. In such a case defects appear to be the effect of non-homogeneous coherent phonon condensation.
Remarkably, these lattice defects exhibit self-similar fractal patterns and provide an example of ``emergence of fractal dislocation structures''~\cite{CrystalFractals} in non-equilibrium dissipative systems. These observations also provide an experimental support to the above mentioned ``theorem'' on the isomorphism between the self-similar properties of fractals\footnote{In some sense self-similarity is considered to be the {\it most important property} of fractals (p. 150 in Ref.~\cite{Peitgen})}~\cite{Peitgen,Bunde} and the algebraic structure of deformed coherent states.
Motivated by such a scenario, our task in this paper is to show that such an isomorphism also exists between the observed phenomenology~\cite{Elia1,Elia2,Elia3} of self-similar, scale free properties of water behavior in the specific conditions below discussed and the formalism of deformed coherent states. Our discussion is limited to a general analysis based on algebraic methods and aimed to account for the self-similarity properties. In a subsequent paper we will present a model on the dynamical molecular behavior of water under the experimental condition specified below. A preliminary, brief description of the model is anticipated in the Appendix B.

The experiments considered in the present paper, published
in~\cite{Elia1,Elia2,Elia3}, have been motivated by those of the Pollack's research group~\cite{Pollack} and are grounded on the studies of the Elia's group in the past couple of decades. We will summarize the measurements in  Section II and for the reader convenience we report full details of the experimental protocol in the Appendix A. The experiments essentially are aimed to the measurements of physically characterizing observables of water samples perturbed by the presence of Nafion, a very hydrophilic polymer, and by filtration. Pollack has indeed shown~\cite{Pollack} laboratory evidence that water in the presence of Nafion acquires singular properties, such as, e.g., impenetrability by impurities in a water stratum of $\approx 200 ~\mu m$ in proximity of the hydrophilic surface of Nafion.
As already mentioned, our discussion is limited essentially to the self-similarity properties which appear from the log-log plots as a laboratory  characterization of all the measurements. Within such limits, without entering the specificity of the molecular dynamics in terms of explicit dynamical modeling, we provide the proof that the observed self-similarity phenomenology is mathematically isomorph to the algebraic structure of deformed coherent states. As well known, the isomorphism between two systems or sets of elements does not mean ``equality'', or analogy in some generic terms, between the two systems. It has a mathematically well defined meaning, so that some difficult to study features or unknown properties of one of the systems can be mapped, if an isomorphism has been found, to well studied features of the other system, for which a sound mathematical formalism has been developed, thus moving from unexplored territories to more familiar ones, sharing the same formal structure with the former ones. A remarkable  example is the Heisenberg discovery that nucleons share the same su(2) algebraic structure of the spin quantum number, e.g. of the electron. This does not mean to propose the interpretation of nucleons in terms of electrons. However, a new branch of physics, i. e. nuclear physics, was born with the discovery of such an isomorphism (and the nucleon isospin was discovered).

Our approach in the discussion presented in Section II is much similar to the one in attempting to catch from the study of symmetry properties as much information as possible, concerning some problem difficult to solve analytically and/or numerically (the extraordinary success of symmetry property studies in high energy physics and condensed matter physics comes here to our minds).
The non-negligible advantage offered by such an approach (in our present case and in physics in general) is that the conclusions one may reach do not depend on specific assumptions or dynamical models. They are of general validity, not to be expected to decay with a too short life-time implied by the sometimes heavy approximations one is forced to introduce in producing a  molecular model or a numerical simulation. For example, in the case of water studies, computational limits put strong constraints in modeling even moderate volumes of liquid water, e.g. $V \approx 100  ~nm^{3}$, by introducing classical limiting assumptions on the water quantum molecular structure~\cite{Wallqvist1999}. The consequent effect is the one of averaging out fluctuations which may turn out in the system collective behavior, such as, e.g., self-similarity~\cite{Guillot2002}. On the other hand, it is of course necessary that in addition to the analysis of the general structure of the phenomenon, an accurate model be formulated by resorting to conventional methods, such as those provided by statistical mechanics and molecular dynamics, able to describe the dynamics at work at a microscopic level.
As a matter of fact, studies along such a direction are on the way and, although the discussion of a dynamical model is out of the scope of the present paper, in the Appendix B we present some preliminary modeling on which we are working.
Section III is devoted to conclusions.

\section{
Phenomenology and algebraic method analysis of nafionized and filtered water
}

Our aim is here limited to analyze by use of algebraic methods the common self-similarity features appearing in
the results of three sets of measurements on water that has been put in contact with Nafion, called Iteratively Nafionized Water (INW), and water that has been iteratively filtered (Iteratively Filtered Water (IFW)). We first describe these experimental results, published in~\cite{Elia1,Elia2,Elia3}, then we present the algebraic method analysis of the common phenomenology of their self-similarity properties.

\begin{figure}
\centering \resizebox{8cm}{!}{\includegraphics{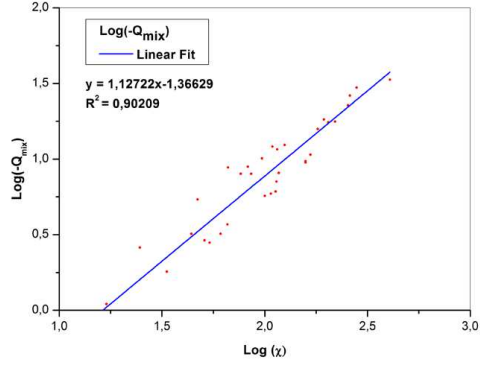}}
\caption{\small \noindent Logarithm of heat of mixing,  $Log (- Q_{mix})$, (with $-Q_{mix} > 0$) of INW with a $NaOH$ solution $0.01 ~m$ ($mol ~kg^{-1}$) as a function of logarithm of the electrical conductivity $\chi$ for INW. Each point represented in the Figure is obtained experimentally measuring the $Q_{mix}$ ($J ~kg^{-1}$) and the electrical conductivity $\chi$, ($\mu S ~cm^{-1}$) of each samples~\cite{Elia1}.
}
\label{fig1}
\end{figure}

From Pollack work~\cite{Pollack} it is known that water in contact of Nafion membranes presents peculiar behavior. We have measured the electrical conductivity $\chi$ of iteratively nafionized water. Nafion membranes of given surface and width are placed in a capsule made either of glass or plastic in contact with $10-20 ~ml$ of pure water. As described in the Appendix A, manual agitation is performed repeatedly so that the liquid laps against the membrane. Then we follow the evolution of $\chi$,  that systematically increases. The procedure is repeated after turning over the membrane. That is iterated for some tens of times, each invariably producing a growth of electrical conductivity. At intervals of a few hours  the membrane is removed from the capsule and left to dry in air. It is then placed back in the nafionized water it came from, and the previous steps (manual agitation, measurement of conductivity, removal of the membrane from the capsule, etc.) are repeated again and again. The measured very  high increase of electrical conductivity $\chi$ (two or about three orders of magnitude) excludes that the phenomenology depends on the impurity release. The impurity release must be rapidly reduced to a null contribute as in a normal washing procedure.
When  the pH and conductivity were measured for samples that had changed their physical-chemical parameters (pH and conductivity) due to aging ($15$ and $30$ days of aging), the measured values lie on the linear trend at the place corresponding to the new coordinates~\cite{Elia1}.  
In Fig. \ref{fig1} we report the plot of the
logarithm of heat of mixing,  $Log (- Q_{mix})$, (with $-Q_{mix} > 0$) of INW with a $NaOH$ solution $0.01 ~m$ ($mol ~kg^{-1}$) as a function of logarithm of the electrical conductivity $\chi$.  The  $pH$ ($-Log [H^+]$) of samples of INW as a function of $Log \chi$ is shown in Fig. \ref{fig2}. All the details on the measurement protocol are described in the Appendix A.

\begin{figure}
\centering \resizebox{8cm}{!}{\includegraphics{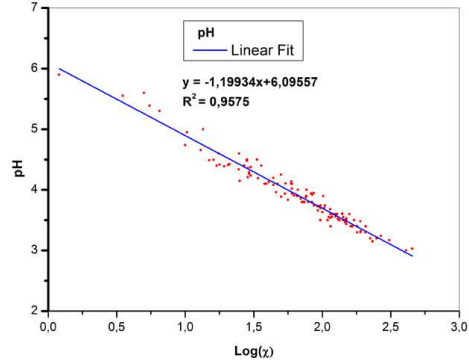}}
\caption{\small \noindent The  $pH$ ($-Log [H^+]$) of $120$ samples of INW as a function of $Log \chi$. Each point represented in the Figure is obtained experimentally measuring the $pH$ and the electrical conductivity $\chi$ ($\mu S ~cm^{-1}$) of each of the $120$ samples~\cite{Elia1}.}
\label{fig2}
\end{figure}


\begin{figure}
\centering \resizebox{8cm}{!}{\includegraphics{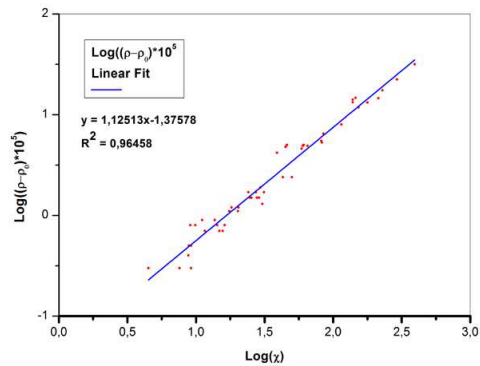}}
\caption
{\small \noindent   Logarithm  of IFW density $(\rho - \rho_{0}) \times 10^5 ~g ~cm^{-1}$, versus  logarithm of electrical conductivity $\chi$ ($\mu S ~cm^{-1}$) for Pirex glass and Millipore filters, irrespective to the number of filtration or  the filter porosity. Each point represented in the Figure is obtained experimentally measuring the density, $\rho$ ($g ~cm^{-1}$) and the electrical conductivity of each samples~\cite{Elia2}.  $\rho_0$ is the density of pure $Milli ~Q$ untreated water.}
\label{fig3}
\end{figure}

Let us now consider the process of iteratively filtering a given volume of pure liquid  water.  The liquid is first
filtered  in vacuum; the resultant filtrate is put through the filtering step again; this filtration is repeated up to $250$ times. It  results~\cite{Elia2,Elia3} that the qualitative effects on water are the same regardless of the filter type, e.g. glass filter, disposable  or ceramic filters. After filtration, electrical conductivity increases by two orders of magnitude, while density shows variations on the fourth decimal digit. Approximately $10-30\% $ of the observed conductivity increases can be attributed to impurities released by the glass filters.
Therefore, we paid careful attention to the impurities released by the glass filters.  We found that the  main chemical impurities are derived from alkaline oxide ($Na_{2}O$) released by the glass. In contact with water, they transform into sodium hydroxide ($NaOH$) and the last substance turns into sodium bicarbonate ($NaHCO_3$) due to atmospheric carbon dioxide ($CO_2$). We therefore systematically determined the sodium concentration of the samples and subtracted the contribution of sodium bicarbonate from the conductivity readings.
The other components of the glass - $SiO_2, ~B_{2}O_3$ and $Al_{2}O_3$ - are very low compared to sodium bicarbonate and they do not contribute significantly to electrical conductivity at the low alkalinity of the water medium, nor do they affect the density, due to their low concentration
(see Tab. 1 of the second quoted paper in Refs.~\cite{Elia3} for their quantitative measurements).
Moreover, the IFW conductivity is not altered by very weak acids, such as  $H_{4}SiO_4$, $H_{3}BO_{3}$, or by $Al_{2}O_3$, deriving from leaching from Pyrex glass filters,  since they are not dissociated in ions in low alkaline solutions, such as those of IFW,  and thus they do not contribute to electrical conductivity~\cite{Elia3}.
We stress that using only samples of pure water there was no possibility for contamination of the filters and no extraneous chemical substances were introduced into the water other than those, mentioned above, deriving from the partial dissolution of the glass solid support. In Fig. 3 it is reported the logarithm  of IFW density $(\rho - \rho_{0})$ versus the logarithm of electrical conductivity $\chi$. $\rho_0$ is the density of pure untreated water.

The fitting by a straight line of the results of the measurements in Figs. \ref{fig1}, \ref{fig2} and \ref{fig3} shows that we are in the presence of a scale free, self-similar phenomenon in the three cases. This result is  reproducible by use of the detailed experimental protocol presented in the Appendix A and to our knowledge it is not described by existing conventional methods of statistical mechanics and molecular modeling. Therefore,  we first analyze the phenomenology  by use of algebraic methods without proposing any  model of the dynamical molecular behavior.   Then, in a future publication, we will present a molecular model. Although this last one is beyond the scope of this paper, nevertheless a preliminary description of a  molecular model on which we are working is anticipated in the Appendix B.

Let us thus proceed in analyzing by algebraic methods the self-similarity (power law) phenomenology of INW and IFW described above (Figs. \ref{fig1}, \ref{fig2} and \ref{fig3}).
The straight line fitting the data in each of the figures \ref{fig1}, \ref{fig2} and \ref{fig3}  is generically represented by the equation
\be \label{sl1}
d = \frac{Log \al}{Log \beta}
\ee
where the ordinate and the abscissa have been denoted by $Log \al$ and $Log \beta$, respectively, with the specific meaning they assume in each of the figures and the reference frame has been translated conveniently,  so that the straight line crosses the zero in each of the three cases (such a  translation is equivalent to divide the ordinate (or multiply the abscissa) by $t$, with $t = 10^c$, where $c$ is the point intercepted by the line on the abscissa axis). It is also understood that the angular coefficient $d$ has the proper $\pm$ sign in each of the $3$ cases.  In the following it is convenient to switch form common logarithms (to base $10$) to the natural ones since the ratio $d$ in Eq.~(\ref{sl1}) does not depend on the chosen base. Eq.~(\ref{sl1}) is equivalent to
\be \label{sl2}
 u_{n,q}(\alpha) \equiv  (q \,\al)^n = 1\,, ~~~~{\rm for ~any}~~ n\,\in \mathbb{N}_+, ~~~{\rm with} ~q \equiv \frac{1}{\beta^d},
\ee
where we have used the notation $(q \,\al)^n \equiv u_{n,q}(\alpha)$, and as customary $\mathbb{N}_+$ denotes positive integers.
The constancy of the angular coefficient $d$ in the plots expresses the scale free character of the relations among physical quantities represented in (each of) the figures and, together with the independence of $n$ of Eq.~(\ref{sl1}), it also expresses their self-similarity properties, namely the ratio $ d = \ln \al \, / \ln \beta$ is independent of the order $n$ of the power to which $\al$ and $\beta$ are {\it simultaneously} elevated. Notice that self-similarity is properly defined only in the $n \rar \infty$ limit.

The above remarks are the ones which are done in the standard analysis of fractal structures~\cite{Bunde}. As already observed in Section I (see footnote 1), self-similarity is considered to be the {\it most important property} of fractals~\cite{Peitgen}. The angular coefficient $d$ in  the figures \ref{fig1}, \ref{fig2} and \ref{fig3} is called the self-similarity dimension, or also the fractal dimension~\cite{Peitgen}.

Now,  the functions $u_{n,q}(\alpha)$
in Eq.~(\ref{sl2}), representing, for any $n  \in \mathbb{N}_+$, the $n$-th power component of the self-similarity relation (the $n$-th stage of the fractal), are readily recognized to be, apart the
normalization factor $1/\sqrt{n!}$, nothing but the restriction to real
$q\,\al$ of the entire analytic functions
in the complex $\al$-plane
\be \lab{un} {\tilde u}_{n,q}(\alpha) = {(q\, \al)^n \over \sqrt{n!}} ~,~\quad
\quad \quad~~ n \in \mathbb{N}_+ ~.
\ee
They form  in the space ${\cal F}$ of the entire analytic functions a basis which is orthonormal under the gaussian measure $d\,\mu (q\,\al) = (1/\pi)e^{-|q\,\al|^2} d\,q\al \,d\,\overline{q\al}$. The factor $1/\sqrt{n!}$
ensures the normalization condition with respect to the gaussian measure.
This means that,  to the extent in which fractals are considered under the point of view of self-similarity, the study of
the fractal properties  may be carried on in the space ${\cal F}$  of the entire analytic functions, by restricting, at the end, the conclusions to real $q\, \al$, $q \,\al \rar {\it Re}(q\,\al)$~\cite{QI2009,NewMat2008}. In other words,  a mathematical isomorphism is recognized to exist between the observed self-similarity properties (Figs. \ref{fig1}, \ref{fig2} and \ref{fig3})  and the deformed coherent states. In order to prove this,
by  closely  following  Ref.~\cite{PLA2012}  we remark %
that ${\cal F}$ is the vector space providing the
representation of the Weyl--Heisenberg
algebra of elements $\{a, a^\dag, 1\}$ with number operator $N = a^{\dagger} a$~\cite{Perelomov:1986tf,Klauder}. ${\cal F}$ is in fact the Fock-Bargmann representation~\cite{Fock} (FBR) of
the (Glauber) coherent states with the identification:
\be \lab{aalpha} N \to \al {d\over d\al} ~ ,~\quad a^\dagger \to
\al ~ ,~\quad a \to
{d\over d \al} ~ .
\ee
In explicit terms, we recall~\cite{NewMat2008,QI2009} that the FBR is the
Hilbert space ${\cal K}$ generated by the basis ${\tilde u}_{n}(\alpha) \equiv {\tilde u}_{n,q}(\alpha)|_{q=1}$, i.e. the
space ${\cal F}$ of entire analytic functions. A one-to-one correspondence
exists between any vector
$\displaystyle{|\psi \rangle}$ in ${\cal K}$ and a function $\psi (\al) \in {\cal F}$. The vector $\displaystyle{|\psi \rangle}$
is then described by the set $\{c_n ; ~c_n\in \mathcal{C},
~\sum_{n=0}^\infty |c_n|^2 = 1 \}$ defined by its expansion in the
complete orthonormal set of eigenkets $\{ |n\rangle \}$ of $N$:
\bea \lab{psi} |\psi \rangle  = \sum_{n=0}^\infty c_n |n\rangle
&\rightarrow&  \psi (\al) = \sum_{n=0}^\infty c_n {\tilde u}_{n}(\alpha), \\
\lab{psi1} \langle \psi|\psi \rangle =  \sum_{n=0}^\infty
|c_n|^2 &=& \int |\psi (\al)|^{2} d \mu (\al) = ||\psi||^{2}=1, \\
\lab{psi2} |n\rangle  &=& \frac{1}{\sqrt{n!}} (a^\dag)^{n}| 0
\rangle ~, \eea
where $|0\rangle$ denotes the vacuum vector, $a |0\rangle = 0$,
$\langle 0|0 \rangle = 1$. The condition $\sum_{n=0}^\infty |c_n|^2 = 1$
(cf. Eq.  (\ref{psi1})) ensures that the series expressing $\psi (\al)$ in Eq.~(\ref{psi}) converges uniformly in any compact domain of the
$\al$-plane, confirming that $\psi (\al)$ is an
entire analytic function, indeed. The explicit expression of
the (Glauber) coherent state  $|\alpha \rangle$ is~\cite{Perelomov:1986tf,Klauder}
\be \lab{(3.2)} |\alpha\rangle = \exp\biggl(-{|\alpha|^2\over 2}
\biggr) \sum_{n=0}^\infty {{\alpha ^n}\over {\sqrt{n!}}} |n\rangle~.
\ee

It is convenient now to put $q = e^\zeta \; ,\; \zeta \in {{\bf
\mathbb{C}}}$. The $q$-deformed algebraic structure
is obtained then by introducing
the finite difference operator ${\cal D}_q$,
also called the $q$-derivative operator. For brevity we do not comment more on this point, see Refs.~\cite{CeleghDeMart:1995,13} for details.  Then, one can show~\cite{CeleghDeMart:1995} that the $q$-deformed coherent state $|q \al \rangle$ is obtained by applying $q^{N}$ to $|\al \rangle$
\be \lab{qN}  q^{N} |\al \rangle = |q \al \rangle =
\exp\biggl(-{{|q\alpha|^2}\over 2}\biggr) \sum_{n=0}^\infty \frac{(q
\alpha)^{n}}{\sqrt{n!}}~ |n\rangle ~. \ee
As observed in Refs.~\cite{PLA2012,NewMat2008,QI2009,Licata2008}, the $n$-th power component $u_{n,q} (\al)$ of the (fractal) self-similarity Eq.~(\ref{sl2})  is ``seen'' by
applying $(a)^{n}$ to $|q \al \rangle$ and restricting to real $q
\al$
\be \lab{nstage}  \langle q \al | (a)^{n} |q \al \rangle =  (q
\alpha)^{n} = u_{n,q} (\al), ~~ \qquad q \al \rar {\it Re} (q \al). \ee
In other words, the operator $(a)^{n}$  acts as a ``magnifying lens''
\cite{Bunde,NewMat2008,QI2009} whose application picks up the $n$-th component of the $q$-deformed coherent state series representing the $n$-th power component $u_{n,q} (\al)$.

Thus, as a result, we have formally established the one-to-one correspondence between the $n$-th power component $u_{n,q} (\al)$ (the $n$-th fractal stage of iteration), with $n =
0,1,2,..,\infty$, and
the $n$-th term in the $q$-deformed coherent state series.
The operator $q^N$ applied to $|\al \rangle$ ``produces'' the
fractal in the functional form of the coherent state $|q \al \rangle$ and therefore it has been called {\it the fractal
operator}~\cite{NewMat2008,QI2009}. Note that $|q \al \rangle$ is actually a squeezed coherent state\cite{CeleghDeMart:1995,Yuen:1976vy} with $\zeta = \ln q $  the
squeezing parameter. Thus, $q^N$ acts in ~${\cal F}$ as the squeezing operator.
The proof of the isomorphism between self-similarity properties of the observed phenomenology and the $q$-deformed algebra of the squeezed coherent
states is formally provided by Eq.~(\ref{qN}) and Eq.~(\ref{nstage})~\cite{PLA2012,NewMat2008,QI2009}. As already said, in a future work we will complement the present result obtained by use of algebraic methods with the formulation of a model of the molecular dynamics.
See the Appendix B for its brief, preliminary presentation.

\section{Conclusions}

The conclusion of our discussion is that an isomorphism exists between the observed scale free, self-similar properties of INW and IFW and the deformed coherent state formalism.
The fractal dimension $d$ has been shown to be related to the $q$-deformation parameter~\cite{PLA2012,NewMat2008,QI2009}, to squeezing and dissipation~\cite{CeleghDeMart:1995}. The relation between  $d$,  $q$ and the squeezing parameter $\zeta$ is given by $- d \ln \beta = \ln q = \zeta$ (cf. Eq.~(\ref{sl2})).

There are however further questions which need to be asked in order to make our analysis more complete and to make it more clear. Among them, the most urgent is perhaps the one related with the identification of the system variables involved in the formation of the coherent states and of their deformation. It might be thus helpful to introduce some clarification for each of the measurements discussed above. Let us start with the case of Fig.~\ref{fig2}. In this case, the system variable involved in coherence is  the (proton) charge density distribution (represented in terms of $pH$) and it is ``deformed'' by its interaction with the Nafion surface\footnote{Indeed Pollack's EZ water is observed~\cite{Pollack} to be arranged in ordered (coherent state) patterns and $pH$ measurements show a peculiar gradient of the proton concentration (orthogonal to the Nafion surface in the geometry of Pollack's experiments~\cite{Pollack}).}.
We may write the (proton) charge density wave function $\sigma ({\bf r}, t)$ as
\be\lab{vs9a}
\sigma ({\bf r}, t) = \sqrt{\rho ({\bf r}, t)}\, e^{i\theta({\bf r}, t)}  ~,
\ee
with real $\rho ({\bf r}, t)$ and $\theta({\bf r}, t)$. The Nafion action is responsible of the spontaneous breakdown of the system $U(1)$ symmetry and non-vanishing $ |\sigma ({\bf r}, t) |^2 = \rho ({\bf r}, t)$  denotes the expectation value of the charge density operator in the system ground state. One may show \cite{Umezawa:1982nv,DifettiBook,DelGiudice:1985} that $\theta({\bf r}, t)$ represents the Nambu-Goldstone (NG) field and that the (space component of the) current is given by
\be \lab{J}
{\bf J}({\bf r}, t) = \frac{1}{m}\rho({\bf r}, t)({\mbox{\boldmath $\nabla$}} \theta ({\bf r}, t) - q {\bf A}({\bf r}, t)).
\ee
where ${\bf A}$ denotes the electromagnetic vector field. On the other hand, the current density is also defined to be proportional to the conductivity $\chi$. This and Eq.~(\ref{J}) gives the relation between $\rho$ and $\chi$. The experiment shows that as an effect of the presence of the Nafion, the conductivity changes. We then assume that $\chi \rar \chi ' = e^{\zeta}\chi$ with $\zeta$ a small ($\zeta < 1$) deformation parameter. It is then easy to show that  the linearity and the coherent state properties lead to the plot of Fig.~\ref{fig2}. The fitting is indeed obtained by tuning the deformation parameter such that  $\zeta = \delta \, \ln \chi$, with $1 + \delta = d$.
The fractal dimension $d$ thus provides a measure of such a dynamical 'deformation', so that the observed  scale free law relating $pH$ and conductivity appears to be the macroscopic manifestations of dissipative local deformations at a microscopic level.

In Fig.~\ref{fig3}, the underlying molecular dynamics of IFW manifests itself in the laboratory observations in terms of the molecular density variations $\rho - \rho_0$ as a function of the electrical conductivity with self-similarity properties.  This means that the filtering process to which water has been undergoing in the described experimental protocol produces molecular rearrangements and displacements. As well known from the study of many-body ordered pattern formation (e.g. as in crystal formation), density behaves as an order parameter and   condensation of long range correlation modes is responsible of the dynamical occurrence coherent states~\cite{Umezawa:1982nv} (for a formal treatment in a specific model see Appendix B and the comments between Eqs.~(\ref{29}) and (\ref{32}) and between Eqs.~(\ref{23}) and (\ref{24a})). By following a derivation similar to the one presented above for Fig.~\ref{fig2}, the variations in the system density (non-homogeneous condensation) are related to the concurrent variations of the conductivity in  the squeezing process characterized by the scale free exponent $d$.

Finally, Fig.~\ref{fig1}  gives for INW the exchanges of mixing heat, $Q_{mix}$, in function of the conductivity at constant pressure and temperature fixed at $25.00 ~{}^{\circ} C \pm 0.001$. Since $T\,\Delta S = \Delta Q$, with
$S$ denoting the entropy, measurement of heat exchange provides the entropy variations in the water molecular configurations as an effect of the nafionization. In such a case, the coherent state is of thermal origin, as it is obtained for example in the thermo field dynamics (TFD) formalism~\cite{Umezawa:1982nv}. We have then a two-mode $SU(1,1)$ coherent state representation related to dissipative (thermal) processes and we can show~\cite{Umezawa:1982nv,DifettiBook,CeleghDeMart:1995} that minimization of the free energy $F$, $dF = 0$, gives, at constant pressure and temperature, $d E = \sum_{k} E_{k} \dot{\cal N}_{k} d t = T d {\cal S}$, where $\dot{\cal N}_{k}$ denotes time derivative of the long range  correlation modes (the Nambu-Goldstone modes). This equation specifically shows that heat exchange  is related to variations of long range correlations out of which  molecular coherent patterns emerge. Thus, again we have coherent state deformations (squeezing). The scale free power law relating heat exchanges and the conductivity changes is then obtained.  Summing up, in the present case the self-similarity properties observed in the experiments reflect the self-similarity properties of the $SU(1,1)$ thermal states.
In this connection, we observe that a  $SU(1,1)$ coherent state representation related to dissipative processes can also be exhibited~\cite{PLA2012} where the notion of topologically non-trivial dissipative phase is introduced and the dynamics is characterized by noncommutative geometry in the plane. Here we omit details on these last issues since they are out of the tasks of this paper. We only remark that a number of specific characterizations of the molecular coherent dynamics can be derived by the above discussed laboratory observations and their understanding in terms of coherent state self-similar properties. Perhaps, of general interest is the possibility, suggested by our approach, of ``extracting reliable information from noisy experiments''~\cite{refer}, such as those described in the present paper. This leads us to one further observation, namely that in the above discussion coherent state formation coexists with non-coherent molecular dynamics, so that one has a two-component system (the coherent and the non-coherent component), with continual migration of molecules from the coherent component to the non-coherent one and vice-versa in an overall stationary regime at fixed temperature. Recent experimental observation points indeed to such a two-component structure of water from ambient temperature to supercooled conditions~\cite{Nature2013} (see also \cite{Bono,AII}).

In order to analyze the self-similarity phenomenology in terms of a specific  molecular dynamics, we need to consider an explicit dynamical model, thus going beyond the limits of the algebraic method analysis to which this paper is devoted.  Preliminary results of the dynamical analysis are presented in the Appendix B.  In our model the molecule dynamics is assumed to be ruled by the interaction of the molecule electrical dipole moment of magnitude $D$ with the radiative electromagnetic (e.m.) field, thus disregarding the static dipole-dipole interaction. The $N$ molecule system is  collectively
described by the properly normalized complex dipole wave field $\chi ({\bf x},t)$ and, by resorting to the analysis of Refs.~\cite{DGV,AII},  we restrict ourselves to the resonant radiative
e.m. modes with $k = \frac{2\pi}{\la} \equiv \om_{0}$. The field
equations  are \cite{Knight,Heitler}:
\bea \nonumber
i \frac{\partial \chi ({\bf x},t)}{\partial t} &=&
\frac{{\bf L}^{2}}{2I} ~\chi ({\bf x},t) - i \sum_{{\bf k}, r}
D~\sqrt{\rho }~\sqrt{\frac{k}{2}} ~ ({\bf \ep}_{r} \cdot {\bf
x}) [u_{r}({\bf k}, t)~ e^{-ikt}
- u_{r}^{\dag}({\bf k}, t)~e^{ikt}]~~\chi ({\bf x},t) ~,\\
 i \frac{\partial u_{r}({\bf k}, t)}{\partial t} &=& i ~D
\sqrt{\rho}~\sqrt{\frac{k}{2}} ~ e^{ikt} \int d \Om ({\bf \ep}_{r}
\cdot {\bf x}) |\chi ({\bf x},t)|^{2}~, \lab{model} \eea
where $u_{r}({\bf k}, t)$
denotes the radiative
e.m. field operator with polarization $r$,
$\rho
\equiv \frac{N}{V}$, $V$ is the volume, and ${\bf \ep}_{r}$ is the polarization vector
of the e.m. mode, for which the condition ${\bf k}
\cdot {\bf \ep}_{r} = 0$ is assumed to hold. We use natural units $\hbar = 1 = c$ and the dipole approximation $\exp(i{\bf
k}\cdot {\bf x}) \approx 1$. We have $\om_{0} \equiv
\frac{1}{I}$, where $I$ denotes the moment of inertia of the molecule; ${\bf L}^{2}$ is the squared
angular momentum operator. The system of $N$ water molecules is  assumed to be spatially homogeneous and in a thermal bath kept at a non-vanishing temperature $T$. For further details on the model see the Appendix B. Preliminary results seem to suggest that the observed self-similarity may occur.

\appendix

\section{The experimental protocol}

For the reader convenience we present here the details of the protocol of three sets of measurements on water under two different
physical treatments~\cite{Elia1,Elia2,Elia3}: water that has been put in contact with Nafion, that we call Iteratively Nafionized Water (INW), and water that has been iteratively filtered (Iteratively Filtered Water (IFW)).

{\it Background and procedure for Iteratively Nafionized Water (INW).}

In order to prepare some water perturbed by the presence of Nafion (Iteratively Nafionized Water (INW))  we followed these steps:

\begin{itemize}
       \item 1. Initially, the pristine membrane is washed $5$ times using $20 ~ml$ of "ultra-pure" ($Milli ~Q^{TM}$) water.
       \item 2. Nafion membranes with a surface of $60-120 ~cm^2$ and a width of $50-180 ~\mu m$, were placed in  a Petri capsule (made either of Pyrex glass or plastic) in contact with $10-20 ~ml$ of $Milli ~Q$ water (electrical conductivity $\chi = 1-2 ~\mu S ~cm^{-1}$). Manual agitation is performed repeatedly so that the liquid laps against the membrane. Then we follow the evolution of $\chi\, (\mu S  ~cm^{-1})$, that systematically increases. The procedure is repeated after turning over the membrane. That is iterated for some tens of times, each invariably producing a growth of electrical conductivity.
       \item 3. At intervals of a few hours (from $3$ to $12$) the membrane is removed from the Petri capsule and left to dry in air ($1-24 ~hours$). It is then placed back in the nafionized water it came from, and step $2$ and $3$ are repeated again.
     \end{itemize}

To obtain a sufficiently high conductivity, i.e. $50-100 ~\mu S ~cm^{-1}$, about $10-20$ iterations of the last two steps are needed. Even though successive iterations invariably determine a growth of electrical conductivity, it has not yet been possible to link quantitatively the number of iterations with the increment of $\chi$. Intuitively, the procedure is akin to a sort of "washing", iterated hundreds of times. The measured  continuous increase of conductivity is such that it is not consistent with the hypothesis of impurity release. In fact the phenomenology takes places independently of the number of steps 2 and 3  of the protocol or of whether the membrane is a pristine one. Membranes used for prolonged periods (months) and for hundreds of steps 2 and 3 behave just like as a new membrane, namely the conductivity always increases at increasing number of steps. It appears, though, that this capability improves with the use of the membrane. In any case if the liquid obtained is completely consumed for experimental measures, and we begin a new procedure using the same membrane with pure water ($10-30 ~ml$) the conductivity increases but does not start at the value obtained in the previous procedure. Because the high number of steps $2$ or $3$ and the very  high increase of electrical conductivity $\chi$ (two or about three orders of magnitude), we can exclude that the phenomenology depends on the impurity release. The impurity release must be rapidly reduced to a null contribute as in a normal washing procedure. Notice that the measures of physical-chemical parameters here reported are obtained after the removal of Nafion membrane from the liquid water. In such a way we obtain information on the effect induced by Nafion on water.

Moreover, we have tested the effect of aging on the samples and we found that when the pH and conductivity are measured for samples that had changed their physical-chemical parameters (pH and conductivity) due to aging over $15$ and $30$ days, the measured values lie on the linear trend at the place corresponding to the new coordinates. The results for samples aged in polyethylene or polypropylene containers are reported in Tab. 1 in Ref.~\cite{Elia1}, from which one can see that after $30$ days from  the sample preparation there is an increase in the
conductivity for the majority of the samples, stability in several of them and a strong decrement in one sample.
It has been observed also that if the aging is performed in presence of small quantity of Nafion membranes, the variation of conductivity with time is practically reduced to zero~\cite{Elia1}. 

{\it Background and procedure for Iteratively Filtered Water (IFW).}

The process of iteratively filtering a given volume ($1-10 ~ml$) of $Milli ~Q$ water consists in:
filtering the liquid in vacuum; taking the resultant filtrate and putting it through the filtering step again; repeating this filtration up to $250$ times.
The following filters were used: $Millipore$ filters made of cellulose nitrate, with porosities of $450$, $200$, $100$ and $25 ~nm$ and $Pyrex$ glass filters having mean porosity of $120$, $65$, $27.5$, $10$ and $2.5 ~\mu\,m$.  One observes~\cite{Elia2,Elia3} that, regardless of the filter type, e.g. $Pyrex$ glass filter (B{\"u}chner), disposable $Millipore$ or $ceramic$ filters, the qualitative effects on water are the same. Upon examining the repeatability of the phenomenon, we decided to first use Pyrex glass filters (B{\"u}chner).
We paid careful attention to the impurities released by the glass filters which might affect electrical conductivity and density. The  main chemical impurities  that we found are derived from alkaline oxide ($Na_{2}O$) released by the glass. In contact with water, they transform into sodium hydroxide ($NaOH$) and the last substance, due to atmospheric carbon dioxide ($CO_2$), turns into sodium bicarbonate ($NaHCO_3$). We therefore systematically determined the sodium concentration of the samples, and subtracted the contribution of sodium bicarbonate from the conductivity readings. The concentrations of impurities deriving from the other components of the glass - $SiO_2, ~B_{2}O_3$ and $Al_{2}O_3$ - are very low compared to sodium bicarbonate
(for numerical values of measured concentrations see Tab. 1 of the second quoted paper in Refs.~\cite{Elia3}).
We stress that the IFW conductivity is not altered by very weak acids such as  $H_{4}SiO_4$, $H_{3}BO_{3}$, or by $Al_{2}O_3$, derived from leaching of Pyrex glass filters, since they are not  dissociated in ions in low alkaline solutions, such as those of IFW,  and thus they do not contribute to electrical conductivity~\cite{Elia3}.

Using Millipore filters for the iterative vacuum filtration process requires use of a sintered glass filter as support. It is observed~\cite{Elia3} that, after filtration, electrical conductivity increases by two orders of magnitude, while density shows variations on the fourth decimal digit. Approximately $10-30\% $ of the observed increases can be attributed to impurities released by the glass filters. At the porosity of Pirex glass filters (R1, mean porosity $120 ~\mu m$) used as a solid support for the Millipore membrane filter, the contribution to electrical conductivity is so slight that its effect does not need to be taken into account.
To exclude the contribution of chemical impurities from inside the Millipore filters, they were rinsed with abundant water until they produced a filtrate with electrical conductivity of $1.2-2.0  ~\mu S ~cm^{-1}$. This procedure is equivalent or better than the sometime suggested soaking. In fact soaking  tends to diminishes the release of impurities, while our goal is to remove all the impurities of electrolyte nature that can increase the electrical conductivity. After the rinsing we can be sure that all the soluble impurities were removed from the filters. We found that the quantitative reproducibility was improved by rinsing the filter with $Milli ~Q$ water after each iterative filtration step, rather than replacing the filter with a new one. It is in fact worth repeating that during the experiment no extraneous chemical substances were introduced into the water other than those deriving from the partial dissolution of the glass solid support. In other words, there was no possibility for contamination of the filters when using only samples of $Milli ~Q$ water.

{\it Electrical conductivity measurements.}
Systematic measurements of specific electrical conductivity were performed on the samples (INW and IFW), using an YSI $3200$ conductometer with an electrical conductivity cell constant of $1.0 ~cm^{-1}$. Before measuring the electrical conductivity of a sample, the cell was calibrated by determining the cell constant $K$ ($cm^{-1}$). The specific conductivity $\chi$ ($\mu S ~cm^{-1}$) was then obtained as the product of the cell constant and the conductivity of the solution. For a given conductivity measuring cell, the cell constant was determined by measuring the conductivity of a $K Cl$ solution having a specific conductivity known to high accuracy, at several concentrations and temperatures. All electrical conductivities were temperature-corrected to $25 ~{}^{\circ} C$, using a pre-stored temperature compensation for pure water.

{\it Calorimetry.}
The  heat of mixing, $Q_{mix}$ ($-Q_{mix} > 0$), of $NaOH$ solution with IFW samples was monitored using a Thermal Activity Monitor (TAM) model $2227$, by Thermometric (Sweden) equipped with a flow mixing vessel. A $P3$ peristaltic pump (by Pharmacia) envoys the solutions (of the solutions of $NaOH$ and of the samples of IFW) into the calorimeter, through Teflon tubes. The flow rates of the two liquids are the same, and are constant in the inlet tubes, so that the solution coming out of the calorimeter has a concentration half the initial one. The mass flow-rate, constant within $1\%$, amounts to $3 \times 10^{-3} g ~s^{-1}$: it was the same for all the experiments. The values of the mixing enthalpies, $\Delta H_{mix}$, were obtained using the following formula:
\be
\Delta H_{mix} \left ({m_{x}}^{i}, ~{m_{y}}^{i}  \rar  {m_{x}}^{f},  ~{m_{y}}^{f} = \frac{dQ}{dt} P_w \right)
\ee
where ($dQ/dt$) is the heat flux ($W$), $P_w$ is the total mass flow-rate of the solvent ($kg ~s^{-1}$) and ${m_{x}}^{i}, ~{m_{y}}^{i}$ and ${m_{x}}^{f}, ~{m_{y}}^{f}$, are the initial and final molalities. $\Delta H_{mix}$ is given in $J ~kg^{-1}$ of solvent in the final solution. For our control, our $Q_{mix}$  represents the difference between the heat of mixing of $NaOH$ solution with the samples of IFW or INW, minus the one with pure untreated $Milli ~Q$  water (heat of dilution of $NaOH$ solution).

{\it Density measurements.}
The solution densities were measured using a vibrating-tube digital density meter (model DMA $5000$ by Anton Paar, Austria) with a precision of $\pm 1 \times 10^{-6} ~g ~cm^{-3}$ and an accuracy of $\pm 5 \times 10^{-6} ~g ~cm^{-3}$. The temperature of the water around the densitometer cell was controlled to $\pm 0.001 ~K$. The densitometer was calibrated periodically with dry air and pure water. As a control for our measurements we use the difference $\rho - \rho_0$ between the density of the samples minus the density of pure untreated $Milli~ Q$ water.

{\it $pH$ measurements.}
The $pH$  were monitored using a $pH$-meter model $micropH ~2002$ by Crison, equipped with a $pH$ electrode for micro-samples, model $5209$.  The electrode specification is: asymmetry potential $< \pm 15 ~mV$, $pH$ sensitivity $4 - 7$ (at $25~ {}^{\circ}C$) $> 98\%$.

{\it Results.}
The results of the measurements of heat of mixing and  $pH$ for INW  and  density for IFW in function of the electrical conductivity are reported in the log-log plots in Figs. 1, 2 and 3, respectively. Their fitting by a straight line shows that we are in the presence of a scale free, self-similar phenomenon in the three cases.

\section{Molecular dynamical model. Preliminary analysis}

We present a preliminary analysis of the water molecular dynamical model described by Eqs.~(\ref{model}).
Since, as observed at the end of Section III, the molecule density is assumed to be spatially uniform, the
only relevant variables are the angular ones. In full
generality, we may expand the field $\chi ({\bf x},t)$ in the unit
sphere in terms of spherical harmonics: $\chi ({\bf x},t) = \sum_{l,m}
\al_{l,m}(t)Y^{m}_{l}(\theta, \phi) $.
By setting $\al_{l,m}(t) = 0$ for $l \neq 0,~ 1$, this reduces
to the expansion in the four levels  $(l,m) = (0,0)$ and $(1,m), m
= 0, \pm 1$. The populations of  these levels are given by $N
|\al_{l,m}(t)|^{2}$ and at thermal equilibrium, in the absence of
interaction, they follow the Boltzmann distribution. The three
levels $(1,m)$, $m = 0, \pm 1$ are in the average equally
populated under normal conditions and we can
safely write $\sum_{m} ~|\al_{1,m}(t)|^{2} =  3 ~|a_{1}(t)|^{2}$, with normalization condition $|\al_{0,0}(t)|^{2} + \sum_{m} ~|\al_{1,m}(t)|^{2}
 = 1$. The  system is invariant under (molecular) dipole rotations, which means that the
amplitude of $\al_{1,m}(t)$ does not depend on $m$, and that the time average of the polarization $P_{{\bf
n}}$  along any direction ${\bf n}$ must vanish in such
conditions.  It is useful to write~\cite{DGV}
\bea \nonumber \al_{0,0}(t) &\equiv& a_{0}(t) \equiv
A_{0}(t)~e^{i\de_{0}(t)} ~,\\
 \al_{1,m}(t) &\equiv&  A_{1}(t)~e^{i\de_{1,m}(t)}~e^{-i\om_{0}t} \equiv a_{1,m}(t)
~e^{-i\om_{0}t} ~,\lab{9}\\
\nonumber u_{m}(t) &\equiv& U (t) e^{i\varphi_{m}(t)}~, 
 \eea
where $ a_{1,m}(t) \equiv A_{1}(t)~e^{i\de_{1,m}(t)}$.
~$A_{0}(t)$, $A_{1}(t)$, $U (t)$,  $\de_{0}(t)$, $\de_{1,m}(t)$ and $\varphi_{m}(t)$  are real
quantities.

Due to the rotational invariance, the rate of change
of the population in each of the levels $(1,m)$, $m = 0, \pm 1$,
equally contributes, in the average,  to the rate of change in the
population of the level $(0,0)$, at each time $t$. In full generality we can set the initial
conditions at $t = 0$ as
\bea \lab{4f} |a_{0}(0)|^{2} &=& \cos^{2} \theta_{0}~,~~
|a_{1}(0)|^{2} = \frac{1}{3}\sin^{2} \theta_{0}~,~~0 < \theta_{0}
< \frac{\pi}{2}~, \\
\lab{4f2} |u(0)|^{2} &=& 0 ~.\eea
By properly tuning the parameter $\theta_{0}$ in its
range of definition one can adequately describe the physical
initial conditions (e.g. $\theta_{0} = \frac{\pi}{3}$
describes the equipartition of the field modes of energy $E(k)$
among the four levels $(0,0)$ and $(1,m)$, $|a_{0}(0)|^{2} \simeq
|a_{1,m}(0)|^{2}, ~m =0, \pm 1$, as typically given by the
Boltzmann distribution when the temperature $T$ is high enough,
$k_{B} T \gg E(k)$). The values zero and $\pi/2$ are excluded since they
correspond to the physically unrealistic conditions for the state
(0,0) of being completely filled or completely empty,
respectively.

From Eqs.~(\ref{model}) one may study  the  ground state of the system for
each of the modes $a_{0}(t)$, $a_{1}(t)$ and $u(t)$. Without reporting the details of the derivation one finds in the mean field approximation~\cite{DGV}:
\bea \lab{27}  \ddot{a}_{0} (t)  &=&
4~{\Om}^{2}\ga_{0}^{2}(\theta_{0}) a_{0}(t) -
4~{\Om}^{2}|a_{0}(t)|^{2}a_{0}(t)~, \\
\lab{31}  \ddot{a}_{1} (t)  &=&  - \si^{2} a_{1}(t) +
12~{\Om}^{2}|a_{1}(t)|^{2}a_{1}(t)~, \\
\lab{18}  \ddot{u} (t)  &=& - \mu^{2}u(t) -
6~{\Om}^{2}|u(t)|^{2}u(t)~, \eea
respectively,
where
$\ga_{0}^{2}(\theta_{0}) \equiv (1/2)(1 + \cos^{2}
\theta_{0})$, $\si^{2} = 2~{\Om}^{2}(1 + \sin^{2} \theta_{0})$ and $\mu^{2} = 2~{\Om}^{2}\cos 2\theta_{0}$, with $\Om \equiv (2D/\sqrt{3}) \sqrt{\rho/2 \, \om_{0}}\, \om_{0}$.
We see that Eq.~(\ref{27}) can be written in the form
\be \lab{28}  
\ddot{a}_{0} (t)  = - \frac{\de }{\de
a_{0}^{*}}V_{0}[a_{0}(t), a_{0}^{*}(t)]~,\ee
where the potential $V_{0}[a_{0}(t), a_{0}^{*}(t)]$ is
\be \lab{28b} V_{0}[a_{0}(t), a_{0}^{*}(t)] = 2 {\Om}^{2}
(|a_{0}(t)|^{2} - \ga_{0}^{2}(\theta_{0}))^{2}
 ~. \ee
Similarly, the
potentials from which the r.h.s. of Eqs.~(\ref{31}) and (\ref{18}) are derivable are
\bea \lab{32}   V_{1}[a_{1}(t), a_{1}^{*}(t)] &=& \si^{2}|a_{1}(t)|^{2} -
6 {\Om}^{2} (|a_{1}(t)|^{2})^{2} ~,\\
\lab{20}  V_{u}[u(t), u^{*}(t)]
&=& 3 {\Om}^{2} (|u(t)|^{2} + \frac{1}{3} \cos
2\theta_{0})^{2}
~,\eea
respectively.
As usual, in order to study the
ground state of the theory, we search for the minima of the
potentials $V$. For  $V_{0}$, let $a_{0,R}(t)$ and $a_{0,I}(t)$ denote the real
and the imaginary component, respectively, of  $a_{0}(t)$: $|a_{0}(t)|^{2} = A_{0}^{2}(t)= a_{0,R}^{2}(t) +
a_{0,I}^{2}(t)$.  One finds a relative maximum of $V_{0}$ at $a_{0} =
0$ and a (continuum) set of minima given by the points on the circle of squared
radius $\ga_{0}^{2}(\theta_{0})$ in the $(a_{0,R}(t),a_{0,I}(t))$
plane:
\be \lab{29}  |a_{0}(t)|^{2} =  \frac{1}{2} (1 + \cos^{2}
\theta_{0}) = \ga_{0}^{2}(\theta_{0}) ~, \ee
We are thus in the familiar case where the cylindrical $SO(2)$ symmetry
(the phase symmetry) around an axis orthogonal to the plane $(a_{0,R}(t),a_{0,I}(t))$
is spontaneously broken. The
points on the circle represent (infinitely many) possible vacua
for the system and they transform into each other under shifts of
the field $\de_{0}$: $\de_{0} \rightarrow \de_{0} + \al$ (SO(2)
rotations in the $(a_{0,R}(t),a_{0,I}(t))$ plane).
The phase
symmetry is broken when one specific ground state is singled out
by fixing the value of the $\de_{0}$ field.
As usual~\cite{ItkZuber}, we
transform to new fields: $A_{0}(t) \rightarrow A_{0}'(t) \equiv
A_{0}(t) - \ga_{0}(\theta_{0})$ and $\de_{0}'(t) \rightarrow
\de_{0}(t)$, so that $A_{0}'(t)  = 0$ in the ground state for
which $A_{0}(t) = \ga_{0}(\theta_{0})$. Use of these new variables
in $V_{0}$ shows that the amplitude $A_{0}'(t)$ describes
a (massive) mode with pulsation $m_{0}
 = 2{\Om} \sqrt{(1 + \cos^{2} \theta_{0})}$ and that
the field $\de_{0}'(t)$ corresponds to a zero-frequency (massless) mode playing the r\^ole of the so-called Nambu-Goldstone
(NG)  collective mode, implied by the spontaneous breakdown
of symmetry. {\it Our assumption is that the perturbations to the water molecular  dynamics induced by iterated interaction with the highly hydrophilic Nafion polymers and the iterated filtration processes are the responsible of the breakdown of  the symmetry described in our dynamical molecular model}.

The value
$a_{0} = 0$, which we have excluded in our initial conditions, cf.
Eq.~(\ref{4f}), on the basis of physical considerations,
consistently appears to be the relative maximum for the potential,
and therefore an instability point out of which the perturbed system
runs away. One can also show that $|u(t)|^{2} = (2/3)~(~|a_{0}(t)|^{2} - \cos^{2} \theta_{0}~)$, which implies that $|u(t)|$ moves away from its vanishing value at $t=0$ (the initial condition Eq.~(\ref{4f2})) as soon as $|a_{0}(t)|$ reaches its minima on the circle or squared radius $\ga_{0}^{2}(\theta_{0})$, considering that   $\theta_{0} \neq 0, \pm \pi, etc.$, as indeed it is since $0< \theta_{0} < \pi/2$.

As well known, {\it the generator of the transformation $\de_{0} \rightarrow \de_{0} + \al$ is the generator of coherent states}~\cite{Perelomov:1986tf,Umezawa:1982nv,DifettiBook}: the infinitely many unitarily inequivalent (i.e. physically inequivalent) vacua, among themselves related by such a transformation, are coherent condensates of the NG modes $\de_{0}$. The family of such coherent states includes squeezed coherent states~\cite{Perelomov:1986tf,Umezawa:1982nv,DifettiBook} parameterized by the $q$-deformation (or squeezing) parameter through the (`form') factors $(q\,\alpha)^n$, for any integer $n$,  thus susceptible to be represented by a straight line in a log-log plot, which is the wanted self-similarity resulting from the (perturbed) dynamical interaction between molecules and radiative e.m. field.

In the case of $V_{1}$, $a_{1} = 0$ is a relative minimum and a
set of relative maxima is on the circle of squared radius
\be \lab{32}  |a_{1}(t)|^{2}  = \frac{1}{6} (1 + \sin^{2}
\theta_{0}) \equiv \ga_{1}^{2}(\theta_{0})  ~. \ee
For $|a_{1}(t)|^{2} =  \ga_{1}^{2}(\theta_{0})$, $U^{2}
= - \frac{1}{3} \cos^{2} \theta_{0} < 0$, which is not acceptable
since  $U$ is real. Thus the amplitude
$A_{1}$ cannot assume the values on the circle of radius
$\ga_{1}(\theta_{0})$, which is consistent with the intrinsic instability of the
excited levels $(1,m)$. One can also show that the conservation
laws in the model (here not reported for brevity) and the reality condition for $U$ require that
$|a_{1}(t)|^{2} \leq \frac{1}{3}\sin^{2}\theta_{0}$ which lies
indeed below $\ga_{1}^{2}(\theta_{0})$, and the value
$\frac{1}{6} \sin^{2} \theta_{0}$ taken by $A^{2}_{1}$ when
$|a_{0}(t)|^{2} = \ga_{0}^{2}(\theta_{0})$ also lies below the
bound. The potential $V_{1}$ thus must be lower than
$\frac{1}{3}\sin^{2}\theta$. These observations show that the consistency between Eqs. (\ref{27}) and
(\ref{31}) is satisfied and
the field $a_{1}(t)$
described by Eq. (\ref{31}) is  a massive field with (real) mass
(pulsation) $\si^{2} =2~{\Om}^{2}(1 + \sin^{2} \theta_{0})$.

For $V_{u}$, we see that $\mu^{2} \geq 0$
for $\theta_{0} \le \frac{\pi}{4}$ and the only minimum is at $u_{0}
= 0$.
This
solution describes the system when the initial condition, Eq.~(\ref{4f2}), holds at any time. However, as
mentioned above this is not
consistent with the dynamical evolution of the system moving away
from the initial conditions exhibited by Eq.~(\ref{27}).  Consistency is  recovered
provided $\theta_{0} > \pi/4$. Then
$\mu^{2} = 2 {\Om}^{2} \cos 2\theta_{0} < 0$ and a
relative maximum of the potential is  at $u_{0} = 0$. A  set of minima is given by the points of the circle of non-vanishing
squared radius $v^{2}(\theta_{0})$ in the $(u_{R}(t),u_{I}(t))$
plane:
\be \lab{23}  |u(t)|^{2}  = - \frac{1}{3} \cos 2\theta_{0} = -
\frac{\mu^{2}}{6{\Om}^{2}} \equiv v^{2}(\theta_{0}) ~, ~~~~
 \theta_{0} > \frac{\pi}{4}~. \ee
These minima represent (infinitely many) possible vacua
for the system and they transform into each other under shifts of
the field $\varphi$: $\varphi \rightarrow \varphi + \al$. The phase symmetry is broken when one specific ground state
is singled out by fixing the value of the $\varphi$ field.
The  fact that  $u_{0} =0$ is now a maximum for the
potential means that the system evolves away from it,
consistently with the similar situation  for the
$a_{0}$ mode. The symmetric solution  at
$u_{0} = 0$ is thus excluded for internal  consistency and the
lower bound $\pi/4$ for $\theta_{0}$ guarantees dynamical self-consistency.

We transform now to new fields: $U(t)
\rightarrow U'(t) \equiv U(t) - v(\theta_{0})$ and $\varphi'(t)
\rightarrow \varphi(t)$ and we find  that $U'(t)$ describes a `massive' mode with real mass
$\sqrt{2|\mu^{2}|} = 2\Om \sqrt{|\cos 2\theta_{0}|}$ (a
quasi-periodic mode), as indeed expected according to the
Anderson-Higgs-Kibble mechanism~\cite{Anderson:1984a,hig,Umezawa:1982nv,DifettiBook}, and that $\varphi'(t)$ is a
zero-frequency mode (a massless mode), also called the
"phason" field \cite{Umez}. $\varphi'(t)$ plays the r\^ole of the  Nambu-Goldstone
(NG) collective mode.
{\it Again, as in the case of the $V_0$ potential,  the generator of the transformation $\varphi(t) \rightarrow \varphi(t) + \al$ is the generator of coherent states}~\cite{Perelomov:1986tf,Umezawa:1982nv,DifettiBook}: the infinitely many unitarily inequivalent  vacua are coherent condensates of the NG modes $\varphi(t)$, whose family includes $q$-deformed (squeezed) coherent states~\cite{Perelomov:1986tf,Umezawa:1982nv,DifettiBook}, also them susceptible to be represented by a straight line in a log-log plot and thus leading us again to the wanted self-similarity resulting from the (perturbed) molecular dynamics.

As a further step, one can show~\cite{DGV} that,  provided $\theta_{0}
> \pi/4$,  which we assume our system is forced to reach under the Nafion and filtering perturbing effects,
\bea \lab{24a}  \dot{U}(t) &=& 2\Om A_{0}(t)A_{1}(t) \cos \al (t) ~,\\
\lab{24b} \dot{\varphi}(t) &=& 2\Om
\frac{A_{0}(t)A_{1}(t)}{U(t)}\sin \al (t) ~, \eea
where $\al \equiv \de_{1}(t) - \de_{0}(t) -
\varphi(t)$. We thus see that $\dot{U}(t) = 0$, i.e. a time-independent
amplitude $\overline{U} = const.$ exists, if and only if the phase
locking relation
\be \lab{25} \al = \de_{1}(t) - \de_{0}(t) - \varphi (t) =
\frac{\pi}{2} \ee
holds. In such a case, $\dot{\varphi}(t) =  \dot{\de_{1}}(t) -
\dot{\de_{0}}(t) = \om$: any change in time of the difference between the
phases of the amplitudes $a_{1}(t)$ and $a_{0}(t)$  is compensated
by the change of the phase of the e.m. field. The phase locking relation (\ref{25}) expresses nothing but the gauge invariance of the theory. Since
$\de_{0}$ and $\varphi$ are the NG modes, eqs. (\ref{25}) also exhibit the coherent feature of the collective
dynamical regime, the ``in phase locked" dynamics of $\de_{0}$ and $\varphi$ coherent condensates, resulting in definitive in the in phase coherence between the system of $N$ dipoles and of the e.m. radiative field.  In such a regime we also have $\overline{A}_{0}^{2} - \overline{A}_{1}^{2} \neq 0$ to be compared with $A_{0}^{2}(t) - A_{1}^{2}(t) \approx 0$ at the
thermal equilibrium in the absence of the collective coherent dynamics.

A final remark concern the finite temperature effects which have not been considered in the above discussion.
Also on such a problem we will focus our study in the planned developments. Here we observe that the $\sqrt{N}$ (appearing in $\sqrt{\rho}$)  in  Eqs.~(\ref{model}) signals strong coupling, namely for large $N$ the interaction time scale is much shorter (by the factor $1/\sqrt{N}$) than typical short range interactions among molecules. Thus for large $N$ the collective interaction is expected to be protected against thermal fluctuations.

Further work is still necessary and some aspects of the model may need much refinement. As said in the text, the discussion of such a model is out of the scope of the present paper. This Appendix is only the anticipation of a preliminary, rudimentary modeling whose final version will be published in a forthcoming paper. There we will also consider the specific spectral analysis on the line of the quantitative results presented in Ref.~\cite{AII}.


\end{document}